\documentclass[aps,pra,twocolumn,10pt,amsmath,amssymb,floatfix,showpacs]{revtex4-1}
\usepackage{graphicx,bm}
\graphicspath{{figs/}}
\renewcommand{\vec}[1]{\bm{#1}}
\newcommand{\abs}[1]{\lvert#1\rvert}
\newcommand{\mat}[1]{\underline{#1}}
\newcommand{\dens}{n}
\newcommand{\magn}{{\vec m}}
\newcommand{\dcur}{J^n}
\newcommand{\curr}{{\vec J}}
\newcommand{\tmat}{{\mathcal T}}
\DeclareMathOperator{\Ei}{Ei}
\DeclareMathOperator{\Li}{Li}

\let\Re\relax\DeclareMathOperator{\Re}{Re}

\begin{document}

\title{Nonlinear spin diffusion and spin rotation in a trapped Fermi gas}
\author{Tilman Enss}
\affiliation{Institut f\"ur Theoretische Physik, Universit\"at
  Heidelberg, 69120 Heidelberg, Germany}
\date{{\today}}
\begin{abstract}
  Transverse spin diffusion in a polarized, interacting Fermi gas
  leads to the Leggett-Rice effect, where the spin current precesses
  around the local magnetization.  With a spin-echo sequence both the
  transverse diffusivity and the spin-rotation parameter $\gamma$ are
  obtained; the sign of $\gamma$ reveals the repulsive or attractive
  character of the effective interaction.  In a trapped Fermi gas the
  spin diffusion equations become nonlinear, and their numerical
  solution exhibits an inhomogeneous spin state even at the spin echo
  time.  While the microscopic diffusivity and $\gamma$ increase at
  weak coupling, their apparent values inferred from the trap-averaged
  magnetization saturate in agreement with a recent experiment for a
  dilute ultracold Fermi gas.
\end{abstract}
\maketitle

\section{Introduction}

Transverse spin diffusion occurs when the magnetization is oriented
along different directions, for instance, in a spin helix.  It has been
observed in spin-echo experiments in helium \cite{hahn1950},
polarized hydrogen, and, recently, ultracold atomic gases
\cite{koschorreck2013, bardon2014, trotzky2015}.  The transverse
magnetization evolves according to a diffusion equation, but there are
additional terms from the precession of the spin currents around the
local magnetization.  This Leggett-Rice effect \cite{leggett1968,
  *leggett1970} is related to the identical spin-rotation effect
\cite{lhuillier1982, *miyake1985} and leads to reactive spin currents
and spin waves, which have been observed in ultracold Fermi gases
\cite{du2008, *du2009, *heinze2013}.  Consider a polarized sample with
small transverse magnetization (small tipping angle).  The transverse
magnetization is conveniently combined into a complex number
$m_+ = m_x+im_y$, which evolves with a complex diffusion coefficient,
\begin{align}
  \label{eq:uniform}
  \frac{\partial m_+}{\partial t}
  = \frac{D_0^\perp}{1-i\mu m_z} \nabla^2 m_+ -i\alpha x_3 m_+,
\end{align}
where $D_0^\perp$ is the transverse diffusivity and $\mu m_z=\gamma$
(at full polarization) the dimensionless spin-rotation parameter.  For
small $\mu\to0$ this is the usual diffusion equation, while for large
$\abs\mu$ the diffusion equation has an imaginary effective
diffusivity and resembles the Schr\"odinger equation \cite{levy1984}.
In a spin-echo pulse sequence, the second term in
Eq.~\eqref{eq:uniform} expresses a linear gradient $\alpha$ of the
external magnetic field along the $x_3$ direction, which winds the
magnetization into a helix.  After a time $t_\pi$, a $\pi$ pulse is
applied, which is equivalent to flipping the sign of $\alpha$.  In the
ensuing time evolution, the helix unwinds until the magnetization is
realigned at the echo time $t_e=2t_\pi$.  In the presence of spin
rotation $\mu\neq0$, realignment at $t_e$ occurs at a phase angle
$\phi \propto \mu$ with respect to the initial orientation at $t=0$
\cite{leggett1968, *leggett1970}, and the value of $\mu$ can be
inferred.  For a strongly interacting Fermi gas the value of $\mu$ has
recently been measured and used to determine the spin-antisymmetric
Fermi-liquid parameter $F_1^a$, while the sign of $\mu$ reveals the
attractive or repulsive character of the effective interaction
\cite{trotzky2015}.  Theoretically, $D_0^\perp$ and $\mu$ for dilute,
homogeneous Fermi gases have been computed using kinetic theory
\cite{jeon1988, *jeon1989, *mullin1992, enss2013trans}.

In an infinite homogeneous system where the only position dependence
arises from the magnetic field gradient $\alpha$, the phase angle
$\phi$ is directly proportional to the microscopic parameter $\mu$.
Instead, for a finite homogeneous box the phase $\phi$, and hence the
apparent value of $\mu$, saturates when the system size is reached
\cite{buu2002}.  Experiments with ultracold atomic gases typically
employ a harmonic trapping potential which is both finite and
inhomogeneous: in this case, the diffusivity $D_0^\perp$, the
Leggett-Rice parameter $\mu$, and the magnetization $m_z$ are strongly
position dependent, and the diffusion equation \eqref{eq:uniform}
becomes nonlinear.  In previous studies the evolution has been
linearized in order to determine the collective mode frequencies and
decay rates in the trapping potential \cite{ragan1995spin,
  *ragan2004leggett, *mullin2006}.  Other studies consider the
nonlinear evolution of the full phase-space distribution for a
nondegenerate or collisionless trapped gas \cite{fuchs2003,
  *piechon2009}.  Here, I numerically solve the nonlinear evolution
equation with position dependent kinetic coefficients including medium
scattering to obtain the transverse magnetization decay and the growth
of the phase $\phi$ for the specific trap geometry used in the experiment.

Kinetic theory \cite{smith1989} is employed to compute the spin
evolution of the trapped gas.  This method is well controlled in the
weak-coupling limit, which is also the parameter regime where
finite-size corrections due to the trapping potential are most
pronounced \cite{trotzky2015}.  On the other hand, at strong coupling
toward unitarity and at low temperature near the superfluid phase
transition of an attractive Fermi gas \cite{bloch2008}, kinetic theory
is expected to receive quantitative corrections from the effects of
pairing and short quasiparticle lifetimes, which are incorporated for
instance when computing transport from the Kubo formula within a
Luttinger-Ward approach \cite{enss2011, *bauer2014, enss2012spin}.

This paper is structured as follows: in Sec.~\ref{sec:kin} the spin
evolution equations for a trapped Fermi gas are derived from kinetic
theory, while Sec.~\ref{sec:evo} compares them to the known
homogeneous limit.  Section~\ref{sec:res} presents the results for the
magnetization profiles and apparent diffusivities, and
Sec.~\ref{sec:con} concludes with a discussion.

\section{Kinetic theory}
\label{sec:kin}

Transport in an interacting Fermi gas may be described by kinetic
theory for quasiparticles, as long as they are sufficiently
long-lived.  For a multi-component Fermi gas with two or more spin
species one has to compute the time evolution of the spin distribution
$\mat n_p$, which is a matrix with components
$\mat n_p=n_{p\sigma\sigma'}(\vec x,t)$ in spin space.  The evolution
equation derived by Landau and Silin \cite{silin1957, baym2008} reads
\begin{multline}
  \label{eq:matkin}
  \frac{\partial\mat n_p}{\partial t}
  + \frac 12 [\nabla_p\mat\varepsilon_p, \nabla_r\mat n_p]_+ 
  - \frac 12 [\nabla_r\mat\varepsilon_p, \nabla_p\mat n_p]_+ \\
  + \frac{i}{\hbar} [\mat\varepsilon_p, \mat n_p]_- 
  = \left( \frac{\partial\mat n_p}{\partial t} \right)_\text{coll}
\end{multline}
where $\mat\varepsilon_p=\varepsilon_{p\sigma\sigma'}(\vec x,t)$ is
the matrix of single-particle energies.  The left-hand side
constitutes the drift term, while the right-hand side describes the
change in the distribution caused by collisions.  Specifically for the
spin-$1/2$ case, $\mat n_p$ and $\mat\varepsilon_p$ are $2\times2$
matrices in spin space, for instance, in the $\uparrow$, $\downarrow$
basis.

The spin matrices can be decomposed in terms of the identity $\mat I$
and the Pauli matrices $\mat{\vec\sigma}$.  The occupation number
matrix is written as
\begin{align}
  \label{eq:matn}
  \mat n_p = \frac12 ( f_p \mat I + \vec\sigma_p \cdot \mat{\vec\sigma} )
\end{align}
where $f_p(\vec x,t)$ is the particle number distribution function and
$\vec\sigma_p(\vec x,t)$ the spin vector distribution in Bloch space.
Similarly, the energy matrix
\begin{align}
  \label{eq:mate}
  \mat\varepsilon_p = \varepsilon_p \mat I + \vec h_p \cdot \mat{\vec\sigma}
\end{align}
combines the spin-independent single-particle energies
$\varepsilon_p(\vec x,t)$ and a magnetic field $\vec h_p(\vec x,t)$.
One may then rewrite Eq.~\eqref{eq:matkin} as
\begin{multline}
  \label{eq:fkin}
  \frac{\partial f_p}{\partial t} + \sum_j
  \left[ \frac{\partial\varepsilon_p}{\partial p_j} \frac{\partial
      f_p}{\partial x_j} - \frac{\partial\varepsilon_p}{\partial x_j}
    \frac{\partial f_p}{\partial p_j} \right. \\
  \left. + \frac{\partial\vec
      h_p}{\partial p_j} \cdot \frac{\partial\vec\sigma_p}{\partial
      x_j} - \frac{\partial\vec h_p}{\partial x_j} \cdot
    \frac{\partial\vec\sigma_p}{\partial p_j} \right]
  = \left( \frac{\partial f_p}{\partial t} \right)_\text{coll}
\end{multline}
and
\begin{multline}
  \label{eq:skin}
  \frac{\partial\vec\sigma_p}{\partial t}
  + \sum_j \left[ \frac{\partial\varepsilon_p}{\partial p_j}
    \frac{\partial\vec\sigma_p}{\partial x_j}
    - \frac{\partial\varepsilon_p}{\partial x_j}
    \frac{\partial\vec\sigma_p}{\partial p_j} \right. \\
  \left. + \frac{\partial \vec h_p}{\partial p_j} 
    \frac{\partial f_p}{\partial x_j}
    - \frac{\partial \vec h_p}{\partial x_j} 
    \frac{\partial f_p}{\partial p_j} \right] 
  - \frac{2}{\hbar} \vec h_p \times \vec\sigma_p
  = \left( \frac{\partial\vec\sigma_p}{\partial t}
  \right)_\text{coll}.
\end{multline}
The spin-rotation term $\vec h_p \times \vec\sigma_p$ is responsible
for the Leggett-Rice effect.  The single-particle energies are
\begin{align}
  \label{eq:speceps}
  \varepsilon_p(\vec x,t)
  & = \frac{p^2}{2m} + V(\vec x), \quad 
    V(\vec x) = \frac m2 \sum_j \omega_j^2 x_j^2
\end{align}
for a Fermi gas in a harmonic trapping potential $V(\vec x)$, which
can be anisotropic with different trapping frequencies $\omega_j$ in
spatial direction $j$.  The magnetic field
\begin{align}
  \label{eq:spech1}
  \vec h_p(\vec x,t) & = -\frac{\hbar}{2} \vec\Omega(\vec x,t), &
  \vec\Omega & = \vec\Omega_0 + \vec\Omega_\text{mf},\\
  \label{eq:spech2}
  \vec\Omega_0(\vec x,t) & = \alpha(t) x_3 \Hat{\vec z}, &
  \vec\Omega_\text{mf} & = \frac{W}{\hbar}\magn(\vec x,t)
\end{align}
is written in terms of the Larmor frequency
$\vec\Omega = \vec\Omega_0 + \vec\Omega_\text{mf}$.  For a spin-echo
protocol it has two contributions: $(i)$
$\vec\Omega_0 = \gamma \vec B(\vec x,t)$ is due to the external
magnetic field $\vec B(\vec x,t)$ with gyromagnetic ratio $\gamma$.  A
spatially constant $\vec B$ is compensated by going to the co-rotating
frame in Bloch space, but a magnetic field $B_z$ gradient of slope
$\alpha$ along the $x_3$ direction winds up the local magnetization
into a spin spiral.  $(ii)$ The second contribution to the Larmor
frequency, $\vec\Omega_\text{mf}$, is a mean-field term proportional
to the local magnetization $\magn$ of a polarized Fermi gas.  It leads
to the precession of the spin current around $\magn$.

The evolution of the full distribution functions $f_p(\vec x,t)$ and
$\vec\sigma_p(\vec x,t)$ is simplified by considering moments with
respect to momentum:
\begin{align}
  \label{eq:moments}
  \dens(\vec x,t) & = \int \frac{d^3p}{(2\pi\hbar)^3}\, f_p(\vec x,t) \\ 
  \dcur_j(\vec x,t) & = \int \frac{d^3p}{(2\pi\hbar)^3}\,
  \frac{\partial\varepsilon_p}{\partial p_j} f_p(\vec x,t) \\
  \magn(\vec x,t) & = \int \frac{d^3p}{(2\pi\hbar)^3}\, 
                               \vec\sigma_p(\vec x,t) \\
  \curr_j(\vec x,t) & = \int \frac{d^3p}{(2\pi\hbar)^3}\,
  \frac{\partial\varepsilon_p}{\partial p_j} \vec\sigma_p(\vec x,t)
\end{align}
with bare velocity $v_{pj}=\partial \varepsilon_p/\partial p_j=p_j/m$.
An additional contribution $f_p \partial \vec h_p/\partial p_j$ to the
spin current is absent for momentum-independent $\vec h_p$.  The local
polarization is defined as
$\vec M(\vec x,t) = \magn(\vec x,t)/\dens(\vec x,t)$ where $\abs{\vec
M(\vec x,t)} \leq 1$.  The
spin current $\curr_j$ is both a vector in Bloch space (bold symbol)
and a vector in position space ($j$ index): it encodes how the
magnetization changes as one goes along the $j$ direction.  The
evolution equations for the moments read, using the specific form of
the single-particle energies \eqref{eq:speceps}--\eqref{eq:spech2},
\begin{align}
  \label{eq:Nevol}
  & \partial_t \dens + \sum_j \nabla_j \dcur_j = 0 \\[-2ex]
  \label{eq:Jnevol}
  & \partial_t \dcur_j
    + \alpha_n \nabla_j \dens + \omega_j^2x_j \dens 
    = -\frac{\dcur_j}{\tau_n} \\
  \label{eq:Mevol}
  & \partial_t \magn
    + \sum_j \nabla_j \curr_j + \vec\Omega_0 \times
    \magn = 0 \\
  \label{eq:Jevol}
  & \partial_t \curr_j
    + \alpha_\parallel P_\parallel \nabla_j \magn
    + \alpha_\perp P_\perp \nabla_j \magn
    + \omega_j^2x_j \magn \notag \\
  & \qquad +\Bigl(\vec\Omega_0 + \frac{W}{\hbar} \magn\Bigr) \times
    \curr_j
   = \left( \frac{\partial\curr_j}{\partial t} \right)_\text{coll}.
\end{align}
The projectors
$P_\parallel \vec a \equiv (\vec a\cdot \Hat{\magn})\Hat{\magn}$ and
$P_\perp \equiv 1-P_\parallel$ give the component of the magnetization
gradient parallel and perpendicular to the local magnetization, respectively.
When deriving these equations, the mean field $\vec\Omega_\text{mf}$
has been retained only in the spin-rotation term and not in the
Poisson brackets in Eqs.~\eqref{eq:fkin} and \eqref{eq:skin}.  This can
be justified as the leading order in a controlled large-$N$ expansion
\cite{enss2012crit}.  In general, the evolution of the currents
depends on the second moments of $f_p$ and $\vec\sigma_p$ which in
turn depend on higher moments.  However, near local equilibrium the
linearized Boltzmann equation relates the higher moments to the lower
ones via the coefficients $\alpha_n$, $\tau_n$,
$\alpha_{\parallel,\perp}$, $\tau_D$, and $W$, and one obtains a
closed set of evolution equations.  These coefficients are discussed
in Sec.~\ref{sec:coe}.

\subsection{Initial conditions}
\label{sec:ini}

\begin{figure}[t]
  \centering
  \includegraphics[width=\linewidth]{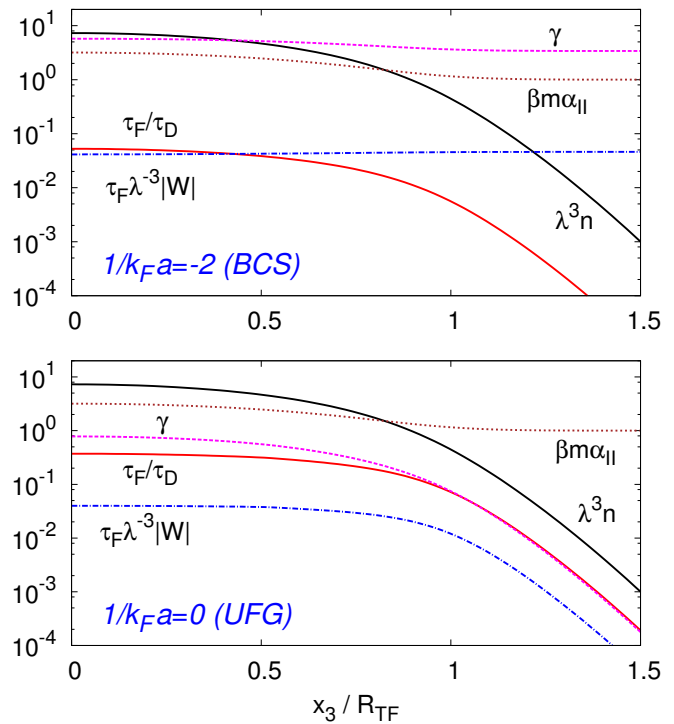}
  \caption{\label{fig:coeffs}(Color online) Kinetic coefficients and
    density profiles in the trap on the BCS side (upper panel) and at
    unitarity (lower panel) at degenerate temperature $(T/T_F)_i=0.2$:
    total phase-space density $\lambda^3n(\vec x)$ (solid black line),
    diffusion parameter $\beta m\alpha_\parallel$ [dotted (brown) line],
    diffusive scattering rate $\tau_F/\tau_D$ [solid (red) line],
    interaction parameter $\tau_F\lambda^{-3}\abs W$ [dash-dotted (blue)
    line], and Leggett-Rice parameter $\gamma=-\tau_DWn/\hbar$ [dashed
    (magenta) line].}
\end{figure}

In the absence of the external magnetic field gradient $\alpha$,
the local equilibrium Fermi distribution is
\begin{align}
  \label{eq:inidist}
  n_{p\pm}(\vec x)
  & = \frac{1}{\exp(\beta(\varepsilon_p(\vec x)-\mu_\pm))+1}
\end{align}
in terms of the chemical potential $\mu_\pm$ of the majority
(minority) component.  The resulting density profile of the fully
polarized gas is
\begin{align}
  \label{eq:inidens}
  \dens(\vec x) & = -\lambda^{-3} \Li_{3/2}(-z_+e^{-\beta V(\vec x)})
\end{align}
with thermal wavelength $\lambda = (2\pi\hbar^2\beta/m)^{1/2}$ and
polylogarithm $\Li_s(z)$.  The fugacity $z_+=\exp(\beta\mu_+)$ of the
majority component is determined by the total particle number $N$ in
the trap of average frequency
$\bar\omega = (\omega_1\omega_2\omega_3)^{1/3}$,
\begin{align}
  \label{eq:totalN}
  N = -\frac{1}{(\beta\hbar\bar\omega)^3} \Li_3(-z_+).
\end{align}
At high temperatures in the Boltzmann regime the density profile is
Gaussian, $\dens(\vec x) \propto \exp(-\beta V(\vec x))$, while at low
temperatures it approaches the Thomas-Fermi profile,
\begin{align}
  \label{eq:densTF}
  \dens(\vec x)
  & = \dens_0 \Bigl( 1-\sum_j \frac{x_j^2}{R_{\text{TF}j}^2} \Bigr)^{3/2}
\end{align}
with Thomas-Fermi radius $R_{\text{TF}j}=(2E_F/m\omega_j^2)^{1/2}$ and
the Fermi energy $E_F = (6N)^{1/3}\hbar\bar\omega$ of a fully
polarized gas.  The density profile is shown in Fig.~\ref{fig:coeffs}
both for weak coupling on the BCS side (upper panel) and for strong
coupling at unitarity (lower panel).  Note that the phase-space
density $\lambda^3\dens$ (solid black line) is unaffected by $s$-wave
scattering in the fully polarized gas.

In the spin-echo protocol \cite{trotzky2015}, the gas is initially
fully polarized, $\abs{\magn}=\dens$ or $\abs{\vec M}=1$, at a tipping
angle $\theta$ from the $z$ axis in Bloch space,
\begin{align}
  \label{eq:inimagn}
  \magn(\vec x,t=0)
  & = (\sin\theta,0,\cos\theta) \dens(\vec x).
\end{align}
This distribution is stationary for $\alpha=0$ and the particle and
spin currents $\dcur_j$ and $\curr_j$ vanish.  When the external
gradient $\alpha$ is switched on along the $x_3$ direction, the
distribution remains independent of $x_1$ and $x_2$, but currents are
generated which can change the distribution along the $x_3$ direction.
Hence, the spin evolution effectively reduces to a one-dimensional
problem for $\magn$ and $\curr_3$ along the gradient direction.

\subsection{Kinetic coefficients}
\label{sec:coe}

The coefficients $\alpha_{\parallel,\perp}(\vec x)$ in
Eq.~\eqref{eq:Jevol} parametrize the strength of the spin current
generated by a magnetization gradient.  Their values are determined by
the Boltzmann equation linearized around the local equilibrium
solution \cite{jeon1988, *jeon1989, *mullin1992, enss2013trans}.
There are different contributions from the longitudinal magnetization
gradient due to the trap potential, and the transverse magnetization
gradient from the helix.  For a fully polarized gas,
\begin{align}
  \label{eq:alphaD}
  \alpha_\parallel(\vec x)
  & = \alpha_n(\vec x) = \frac{\dens(\vec x)}{m\chi(\vec x)} \\
  \alpha_\perp(\vec x) & = \frac{P(\vec x)}{m\dens(\vec x)}
\end{align}
where
$\chi(\vec x) = -\lambda^{-3}\beta \Li_{1/2} \bigl(-z_+e^{-\beta
  V(\vec x)}\bigr)$
is the local susceptibility, and
$P(\vec x) = -\lambda^{-3}\beta^{-1} \Li_{5/2} \bigl(-z_+e^{-\beta
  V(\vec x)}\bigr)$
the local pressure of a free Fermi gas.  In Fig.~\ref{fig:coeffs},
$\alpha_\parallel$ is plotted as the dotted line: for a high temperature
or low density in the outer regions of the trap it reaches the
Boltzmann limit $\alpha_\parallel=\alpha_\perp=1/\beta m$, while for
a high density at the trap center it is enhanced and approaches the
low-temperature limit $\alpha_\parallel=v_F^2/3$ \cite{leggett1968,
  *leggett1970}.  In general, $\alpha_\parallel$ and $\alpha_\perp$
acquire interaction corrections \cite{jeon1988, *jeon1989,
  *mullin1992}, but for large initial polarization and $s$-wave
scattering these are negligible.

Second, the coefficients $\tau_\parallel(\vec x)$ and
$\tau_\perp(\vec x)$ parametrize the decay of the spin current due to
scattering \cite{jeon1988, *jeon1989, *mullin1992},
\begin{multline}
  \label{eq:sigmacoll}
  \left( \frac{\partial\curr_j}{\partial t} \right)_\text{coll}
  = \int \frac{d^3p}{(2\pi\hbar)^3}\, \frac{\partial \varepsilon_p}{\partial p_j}
  \left( \frac{\partial\vec\sigma_p}{\partial t} \right)_\text{coll} \\
  = -\frac{P_\parallel \curr_j}{\tau_\parallel} 
  -\frac{P_\perp \curr_j}{\tau_\perp}.
\end{multline}
The component of the current $P_\parallel \curr_j$ parallel to the
local magnetization $\magn$ decays with the longitudinal diffusion
time $\tau_\parallel$, while the transverse component
$P_\perp \curr_j = (1-P_\parallel) \curr_j$ decays with the transverse
diffusion time $\tau_\perp$.  Both differ for a polarized, strongly
degenerate Fermi gas \cite{meyerovich1985, jeon1988, *jeon1989,
  *mullin1992, enss2013trans}: all the states between the majority and
the minority Fermi surfaces of a polarized gas are available for
transverse scattering and $\tau_\perp$ can be much lower than
$\tau_\parallel$.  For the experiment at hand \cite{trotzky2015},
however, the lowest temperatures reached at the trap center are around
$T/T_F\gtrsim 0.3$, and previous studies have shown that in this
temperature range $\tau_\parallel$ and $\tau_\perp$ are nearly equal
\cite{enss2013trans}.  It is therefore justified to work with a single
decay time $\tau_D = \tau_\perp \approx \tau_\parallel$
\cite{enss2013trans},
\begin{align}
  \frac{\hbar}{\tau_\perp}
  & = \frac{\sinh(\beta h)}{C_\perp}
  \frac{1}{(2\pi\hbar)^8} \int d^3p_1\dotsc d^3p_4\notag\\
  & \times \delta^{(3)}(\vec p_1+\vec p_2-\vec p_3-\vec p_4)\,
  \delta(\varepsilon_{p_1} + \varepsilon_{p_2} - \varepsilon_{p_3} -
  \varepsilon_{p_4}) \notag\\
  & \times \abs{\tmat(\vec p_1,\vec p_2)}^2\,
  [e^{-\beta h} n_{1+} n_{2+} + e^{\beta h} n_{1-} n_{2-}] \notag\\
  \label{eq:tauperpgen}
  & \times (1-n_{3+})(1-n_{4-}) v_{1j} (v_{1j} - v_{2j})
\end{align}
with magnetic field $h=(\mu_+-\mu_-)/2$ and normalization constant
\begin{align}
  \label{eq:CD}
  C_\perp = \int \frac{d^3p}{(2\pi\hbar)^3} \, v_{pj}^2 (n_{p+}-n_{p-}).
\end{align}
The $T$ matrix $\tmat(\vec p_1,\vec p_2)$ describes $s$-wave
scattering between particle $(\vec p_1,+)$ and particle $(\vec p_2,-)$.  In
order to derive explicit expressions, the $T$ matrix for ultracold
fermions with $s$-wave contact interactions has to be used.  The
two-body $T$ matrix
\begin{align}
  \label{eq:tmat0}
  \tmat_0(\vec p_1,\vec p_2) = \frac{4\pi\hbar^2}{m}\, \frac{1}{a^{-1} +ik}
\end{align}
is given in terms of the relative wavenumber
$k=\abs{\vec p_1-\vec p_2}/2\hbar$ and the $s$-wave scattering length
$a$.  At weak coupling $\abs a\to0$,
$\tmat_0(\vec p_1,\vec p_2) \to 4\pi\hbar^2a/m$ is the regularized
bare contact interaction.  The BCS-BEC crossover goes from the weakly
attractive BCS regime ($1/k_Fa \lesssim -1$) via the unitary Fermi gas
(UFG, $1/k_Fa=0$) to the repulsive fermion branch ($1/k_Fa \gtrsim 1$)
above the BEC ground state \cite{bloch2008}.  At
strong coupling the many-body $T$ matrix is needed, which includes
medium scattering.  In the ladder approximation, which is the leading
order of the large-$N$ expansion in the number of fermion flavors
\cite{nikolic2007, enss2012crit}, the full $T$ matrix reads
\begin{multline}
  \label{eq:tmat}
  \tmat^{-1}(\vec p_1,\vec p_2)
  = \tmat_0^{-1}(\vec p_1,\vec p_2)\\
  + \int \frac{d^3p}{(2\pi\hbar)^3}
  \frac{n_{\vec p,+} + n_{\vec p+\vec p_1+\vec p_2,-}}
  {\varepsilon_{\vec p_1} + \varepsilon_{\vec p_2} - \varepsilon_{\vec
      p} - \varepsilon_{\vec p+\vec p_1+\vec p_2}+i0}.
\end{multline}

The $T$ matrix is computed numerically, and Fig.~\ref{fig:coeffs}
shows the resulting spin diffusion rate $\tau_F/\tau_D$ in units of
the Fermi frequency $1/\tau_F=E_F/\hbar$ as the solid (red) line: it is
highest in the trap center and decreases proportional to the density
in the outer regions.  At unitarity, the scattering rate is about $10$
times larger than at weak coupling $1/k_Fa=-2$.  In the Boltzmann
regime,
$\hbar/\tau_D = \frac{4\sqrt 2 n \lambda^3}{3\pi\beta} \bigl[ 1 -
\beta\varepsilon_B - (\beta\varepsilon_B)^2 \exp(\beta\varepsilon_B)
\Ei(-\beta\varepsilon_B) \bigr]$
where $\Ei(x)$ is the exponential integral and
$\varepsilon_B=\hbar^2/ma^2$ \cite{enss2013trans}.  At unitarity the
scattering cross section decreases with temperature, and
$\hbar/\tau_D = 4\sqrt 2 n \lambda^3/(3\pi\beta)$ \cite{bruun2011,
  *bruun2011spin}.

Third, the mean field $\vec\Omega_\text{mf} = W \magn(\vec x,t)/\hbar$
describes the precession of the spin current around the local
magnetization and is given by a momentum average of the \emph{real
  part} of the many-body $T$ matrix $\tmat(\vec p_1,\vec p_2)$ over
the momentum states between the majority and the minority Fermi surfaces,
weighted by the velocity squared \cite{enss2013trans},
\begin{multline}
  \label{eq:spinrot}
  W = \frac{1}{C_\perp \abs\magn} \int
  \frac{d^3p_1}{(2\pi\hbar)^3} \frac{d^3p_2}{(2\pi\hbar)^3} 
  v_{1j} (v_{1j}-v_{2j}) (n_{1+}-n_{1-}) \\
  \times (n_{2+}-n_{2-}) \Re \tmat(\vec p_1,\vec p_2).
\end{multline}
At weak coupling $W=\tmat_0(0,0) = 4\pi\hbar^2a/m$ agrees with the
bare interaction, which is real.  At unitarity $1/a\to0$, $\tmat_0$
becomes purely imaginary and $W$ appears to vanish along with
$\Re\tmat_0$.  This is indeed observed at high temperatures, but at low
temperatures the \emph{many-body} $T$ matrix $\tmat$ acquires a real
part due to medium scattering, and $W\neq 0$ \cite{trotzky2015}.  This
is shown as the dash-dotted (blue) curve in Fig.~\ref{fig:coeffs}: for
weak coupling, $W\approx 4\pi\hbar^2a/m$ is constant independent of the
density and position in the trap, while at unitarity it decreases with
density in the outer regions of the trap.  Note that the very similar
values for $W$ at the trap center are coincidental: at weak coupling
$W$ is given essentially by the bare coupling, while at unitarity it
is purely a many-body effect.

Spin rotation is characterized by the dimensionless Leggett-Rice
parameter
\begin{align}
  \label{eq:gamma}
  \gamma = \mu n
  = -\frac{\tau_DWn}{\hbar}
\end{align}
which is plotted as the dashed (magenta) line in Fig.~\ref{fig:coeffs}.
At weak coupling,
\begin{align}
  \label{eq:gammaweak}
  \tau_D & \propto \frac{1}{a^2n^{4/3}}, & W & \propto a, &
  \gamma & \propto -\frac{1}{an^{1/3}},
\end{align}
hence $\gamma$ becomes large and only weakly dependent on the density.  At
unitarity, on the other hand, $\gamma$ is a purely many-body effect,
much smaller, and roughly proportional to the density.

\section{Analytical solutions in limiting cases}
\label{sec:evo}

The evolution equations \eqref{eq:Nevol}--\eqref{eq:Jevol} leave the
density profile largely invariant, but the spin distribution changes
dramatically as the magnetization is wound up into a helix according
to the equations
\begin{align}
  \label{eq:contin}
  & \partial_t \magn + \sum_j \nabla_j \vec J_j + \alpha x_3
    \Hat{\vec z} \times \magn = 0 \\
  \label{eq:jevol}
  & \partial_t \vec J_j + \alpha_\parallel P_\parallel \nabla_j \magn
    + \alpha_\perp P_\perp \nabla_j \magn + \omega_j^2x_j \magn \notag\\
  & \qquad\qquad + (\alpha x_3 \Hat{\vec z} + \frac{W}{\hbar} \magn) \times
    \vec J_j = -\frac{\vec J_j}{\tau_D}.
\end{align}
The full numerical solution of these equations is presented
in Sec.~\ref{sec:res}.  In order to gain a qualitative understanding
of the spin evolution in a trapped gas, it is instructive to consider
first the approximate analytical solutions in the homogeneous case.

If the scattering time $\tau_D$ is much shorter than any other
relevant time scale, for instance, the dephasing time
$\tau=(D^\perp\alpha^2)^{-1/3}$ due to the helix, the current reaches
a steady state and its time derivative vanishes in the rotating frame,
$\partial_t \vec J_j = -\alpha x_3 \Hat{\vec z} \times \vec J_j$.
Defining $D_0^\perp = \alpha_\perp \tau_D$ and
$\mu \magn = -\tau_D \vec\Omega_\text{mf} = \gamma \vec M$ one finds
\begin{align}
  \vec J_j + D_0^\perp \nabla_j \magn + \tau_D \omega_j^2 x_j \magn 
  - \mu \magn \times \vec J_j = 0.
\end{align}
This equation is solved by the steady-state current
\begin{multline}
  \label{eq:steady}
  \vec J_j = -\frac{D_0^\perp}{1+\mu^2\magn^2} \Bigl\{ \nabla_j
    \magn + \mu \magn \times \nabla_j \magn \\
    + \mu \magn(\mu \magn \cdot \nabla_j \magn)\Bigr\}
    - \tau_D \omega_j^2 x_j \magn
\end{multline}
where $D_0^\perp$, $\mu$, and $\tau_D$ may still depend on position;
the last term arises due to the trapping potential.  Inserting this
current into the continuity equation for the magnetization
\eqref{eq:contin} yields
\begin{multline*}
  \partial_t \magn = -\alpha x_3 \Hat{\vec z} \times \magn
  + \sum_j \omega_j^2 \nabla_j (\tau_D x_j \magn) \\[-2ex]
  + \nabla_j \frac{D_0^\perp}{1+\mu^2\magn^2} \Bigl\{ \nabla_j
  \magn + \mu \magn \times \nabla_j \magn 
  + \mu \magn(\mu \magn \cdot \nabla_j \magn)\Bigr\}.
\end{multline*}
The Leggett solution \cite{leggett1968, *leggett1970} is recovered in
the homogeneous limit $\omega_j=0$, where $\magn^2$ remains constant
in space:
\begin{align*}
  \partial_t \magn = -\alpha x_3 \Hat{\vec z} \times \magn
  + \frac{D_0^\perp}{1+\mu^2\magn^2} \bigl\{ \nabla^2
  \magn + \mu \magn \times \nabla^2 \magn \bigr\}.
\end{align*}
In this case the longitudinal magnetization $m_z$ remains unchanged,
while the transverse magnetization $m_+ = m_x+im_y$ evolves as
\begin{align}
  \label{eq:evolhom}
  \partial_t m_+ & = -i\alpha x_3 m_+ +
  D_\text{eff}^\perp(1+i\mu m_z) \nabla^2 m_+
\end{align}
with effective diffusivity
$D_\text{eff}^\perp = D_0^\perp/(1+\mu^2\magn^2)$.  Since $\magn^2$ is
constant in space, this is now a \emph{linear} diffusion equation,
albeit with a complex diffusion coefficient.

In the spin-echo protocol, the gradient $\alpha$ winds up the
transverse magnetization $m_+ = m_x+im_y$ into a helix along the $x_3$
direction.  A transverse spin current $J_{3+} \sim \partial_3 m_+$
appears, which tends to smooth the helix.  If $J_{3+}$ has a component
perpendicular to the local magnetization
$\magn \approx m_z\Hat{\vec z}$ (at small tipping angle) it precesses
around it at frequency $\Omega_\text{mf}=Wm_z/\hbar$.  At time
$t_\pi$, a $\pi$ pulse around the $y$ axis in Bloch space is applied;
this is equivalent to reversing the sign of $\alpha$.  The subsequent
time evolution unwinds the helix until the echo time $t_e=2t_\pi$,
where the transverse magnetization is again homogeneous.  In the
absence of spin rotation, $\gamma=0$, the modulus of the transverse
magnetization $A(t)=\abs{m_+(t)}$ at time $t_e$ decays as a cubic
exponential \cite{leggett1968, *leggett1970}
\begin{align}
  \label{eq:expdecay}
  A(t_e) 
  = A_0 \exp\Bigl(-\frac{D_0^\perp\alpha^2t_e^3}{12}\Bigr)
\end{align}
where $A_0=A(0)$.  This result is approximately correct even for
$\gamma\neq0$ if the tipping angle $\theta$ is small,
$\abs{m_+}\ll \abs{m_z}$, and for short times.  For finite $\gamma$,
the magnetization decay is slowed down and is given by
\cite{trotzky2015}
\begin{align}
\label{eq:lambert}
  A(t_e) & = A_0 \sqrt{\frac{1}{\eta}\, 
  \mathcal{W}\left(\eta \exp{\left[\eta-\frac{D_0^\perp \alpha^2
           t_e^3}{6(1 + \gamma^2 M_z^2)}\right]}\right)}\,,\\
\label{eq:phievol}
  \phi(t_e) &= \gamma M_z \ln \left(\frac{A(t_e)}{A_0}\right)\,,
\end{align}
where $\eta = \gamma^2(A_0/n)^2 / (1 + \gamma^2 M_z^2)$ for
polarization $M_z$, and $\mathcal{W}(z)$ is the Lambert-$W$ function.
This solution for the homogeneous system is used in the analysis of
the experimental data to fit the diffusivity $D_0^\perp$ and the
apparent Leggett-Rice parameter $\gamma$ from the measured
magnetization decay and phase shift $\phi$ of the trapped system.  In
the following the full spin evolution in the trap is computed
explicitly to determine to what extent the homogeneous solution is
still applicable to the trapped gas.

\section{Results}
\label{sec:res}

In the trapped gas, the assumptions which led to the analytical
solutions in Sec.~\ref{sec:evo} (steady-state current, homogeneity)
are not satisfied.  Instead, a full numerical solution of the spin
evolution equations, \eqref{eq:contin} and \eqref{eq:jevol}, is necessary.
The initial condition in the experimental protocol \cite{trotzky2015}
is a fully polarized cloud of fermionic atoms with a thermal profile,
\eqref{eq:inidens} and \eqref{eq:inimagn}, for tipping angle $\theta$.
This distribution is stationary in the absence of an external magnetic
field gradient $\alpha$.  Due to the density profile,
\eqref{eq:inidens}, also the coefficients $\alpha_\perp(\vec x)$,
$\tau_\perp(\vec x)$, and $W(\vec x)$ depend on the position in the
trap.  For a small tipping angle $\abs{m_+} \ll \abs{m_z}$ the gas
remains almost fully polarized, hence the density profile and the
coefficients are time independent even in the presence of the gradient
$\alpha$.

In the numerical solution, the experimental parameters
\cite{trotzky2015} are used: gradient $\alpha=1.67\mu$m$^{-1}$kHz,
trap frequency $\omega_3 = 2\pi\times 750$\,Hz,
$\Bar\omega=2\pi\times 470$\,Hz, roughly $N\sim40000$ atoms of $^{40}$K
with a Fermi energy of $E_F=2\pi\hbar\times 29$\,kHz, and Thomas-Fermi
radius $R_{TF}=5.1\mu$m.  Even though the Fermi gas is initially fully
polarized, the Fermi wave vector $k_F=12\mu$m$^{-1}$ and the Fermi time
$\tau_F=0.0087$\,ms are taken for a hypothetical balanced gas of the
same atom number in order to recover the standard relation
$n=k_F^3/3\pi^2$ for the total density \cite{trotzky2015}.  At the
lowest experimental temperature, $T\approx 280$\,nK, the initial reduced
temperature is $(T/T_F)_i=0.2$ and $\lambda=0.52\mu$m.  The
phase-space density at the trap center reaches $\lambda^3n=7.3$, well
within the quantum degenerate regime.

\begin{figure}[t]
  \centering
  \includegraphics[width=\linewidth]{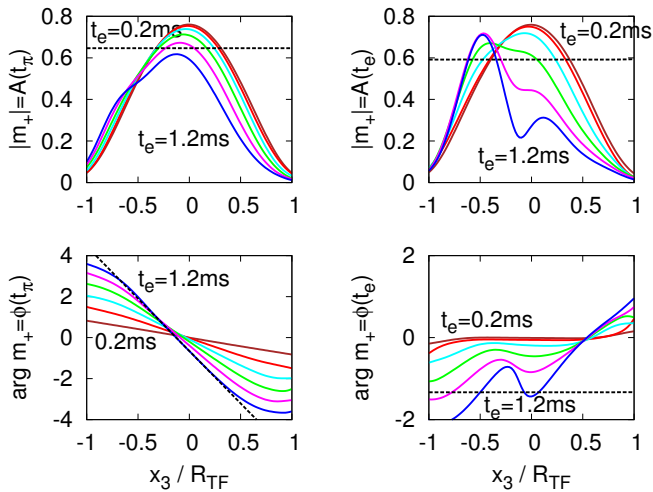}
  \caption{\label{fig:profiles}(Color online) Transverse magnetization
    profiles on the BCS side $1/k_Fa=-2$ at temperature
    $(T/T_F)_i=0.2$ and times $t_\pi$ just before the $\pi$ pulse
    (left column), and at the echo time $t_e=2t_\pi$ (right column).
    Upper panels display the amplitude $A=\abs{m_+}$; lower
    panels, the phase angle $\phi=\arg m_+$.  Solid curves in each
    panel are for a trapped gas at $t_e=0.2$\,ms to $t_e=1.2$\,ms, in
    comparison with a homogeneous gas at $t_e=1.2$\,ms (dashed line).}
\end{figure}
In Fig.~\ref{fig:profiles}, the helical state of the local
magnetization in the trap at $1/k_Fa=-2$ is shown for different times
$t_\pi$ (left column).  At this time, the $\pi$ pulse is applied
(equivalently, the sign of $\alpha$ is reversed), and the subsequent
time evolution unwinds the helix until the echo time $t_e=2t_\pi$
(right column), which is chosen between $t_e=0.2$\,ms and $t_e=1.2$\,ms.
For comparison, the density profiles for a homogeneous gas with
$t_e=1.2$\,ms are shown as dashed lines.  At time $t_\pi$, the $\phi$
profiles of the trapped and the homogeneous gas have a similar slope and
are shifted toward negative phase angles by spin rotation.  At time
$t_e$, instead, the phase angle of the homogeneous gas is again
homogeneous, but $m_+$ for the trapped gas remains in a helical state.
Therefore, the trap-averaged transverse magnetization decays quickly
even though the local $m_+$ is still sizable, and the apparent slope
$\phi(t_e)/M_z\ln(A(t_e)/A_0)$ is lower than in the homogeneous case,
where it reaches the microscopic value $\gamma = -Wm_z\tau_D/\hbar$
according to Eq.~\eqref{eq:phievol}.  As we see below, this
effect leads to a saturation of the apparent $\gamma$ for weak
coupling.  

Note that the magnetization profiles are not symmetric with respect to
$x_3$: this is due to the spin-rotation term in Eq.~\eqref{eq:evolhom}
producing an additional phase shift on top of the gradient $\alpha$.
Consequently, if one reverses the sign of the scattering length $a$,
the sign of $\gamma$ is approximately reversed, and the resulting
magnetization profile is the mirror image with $x_3 \mapsto -x_3$.
Note also that at strong coupling $1/k_Fa=0$, the phase profile at
$t_e$ is nearly homogeneous at the trap center (not shown), and the
kinetic coefficients for the trapped gas approach those in the
homogeneous case.

\begin{figure}[t]
  \centering
  \includegraphics[width=\linewidth]{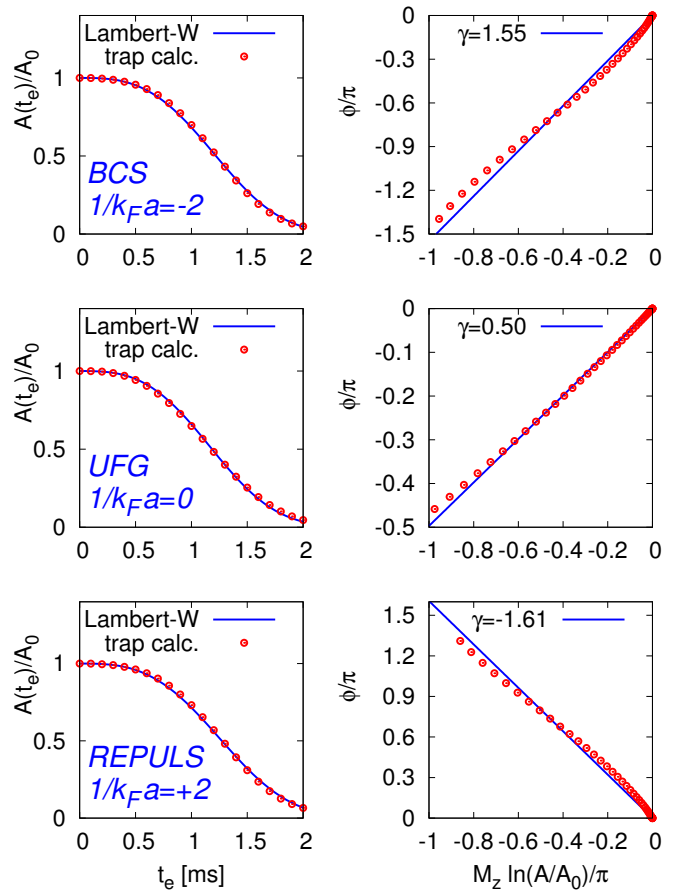}
  \caption{\label{fig:ampphase}(Color online) Amplitude and phase of
    the transverse magnetization $m_+$ at temperature $(T/T_F)_i=0.2$
    and small tipping angle $\theta=\pi/30$ ($M_z=0.995$).  Left:
    analytical solution, \eqref{eq:lambert} (solid line), with
    $D_0^\perp$ adjusted to fit the trap calculation (circles).
    Right: fit of $\gamma$ from the slope of the phase shift $\phi$ in
    Eq.~\eqref{eq:phievol}.  Top, $1/k_Fa=-2$ (BCS side); center,
    $1/k_Fa=0$ (unitarity); bottom, $1/k_Fa=2$ (repulsive branch).}
\end{figure}
Figure~\ref{fig:ampphase} shows the decay of the trap-averaged
transverse magnetization with time (left column), and the growth of
the phase angle $\phi$ with the slope of the magnetization decay
(right column).  From these plots, the apparent Leggett-Rice parameter
$\gamma = \phi(t_e)/M_z\ln(A(t_e)/A_0)$ can be read off as the slope
of the curve through the origin.  This value of $\gamma$ is then used
to fit the magnetization decay on the left side of
Fig.~\ref{fig:ampphase} to the analytical decay function
\eqref{eq:lambert} with diffusivity $D_0^\perp$.  Here, I follow the
analysis of the experimental data in Ref.~\onlinecite{trotzky2015} and
fit the decay of the trapped gas to the homogeneous solution.

\begin{figure}[t]
  \centering
  \includegraphics[width=\linewidth]{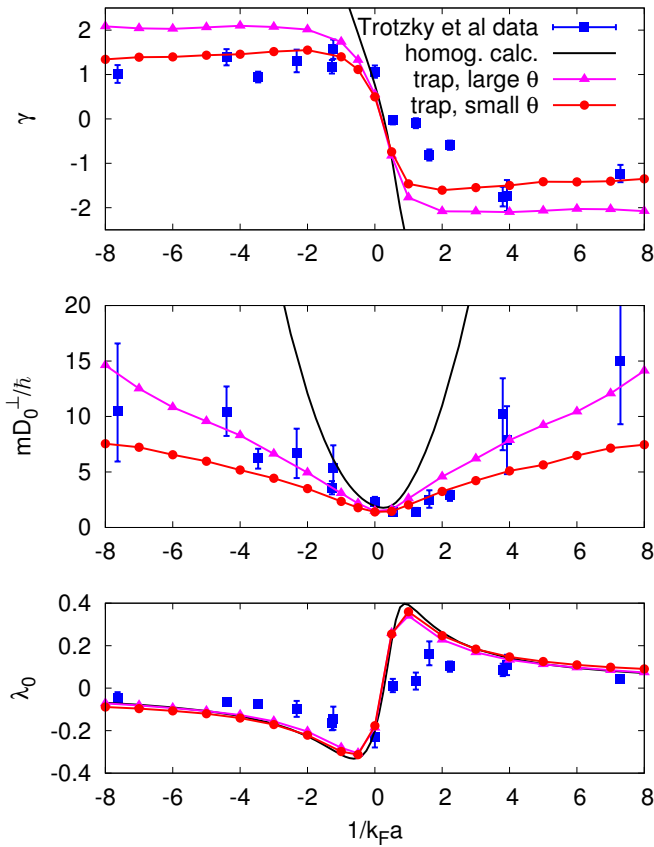}
  \caption{\label{fig:intdiff}(Color online) Interaction effect on
    spin transport: Leggett-Rice parameter $\gamma$, bare diffusivity
    $D_0^\perp$, and interaction parameter $\lambda_0$ as a function
    of $1/k_Fa$ at degenerate temperature $(T/T_F)_i=0.2$.  Results
    for small ($\theta=\pi/30$; circles) and large ($\theta=\pi/3$;
    triangles) tipping angles are shown and compared to the
    experimental data \cite{trotzky2015} (squares).}
\end{figure}
In Fig.~\ref{fig:intdiff}, the apparent Leggett-Rice parameter
$\gamma$ extracted above is plotted versus the interaction strength, from
the BCS regime (left) via unitarity ($1/k_Fa=0$) to the repulsive
branch (right).  For a trapped gas, I find that the apparent $\gamma$
saturates for weak coupling on both the BCS and the repulsive sides.  This
behavior differs qualitatively from the homogeneous case, where
$\gamma$ continues to increase linearly at weak coupling
[cf.\ Eq.~\eqref{eq:gammaweak}] (solid black line).  The trap calculation
thus explains the saturation of $\gamma$ observed in experiments
\cite{trotzky2015}.  The absolute value of $\gamma$ for the trapped
gas depends on the tipping angle $\theta$ of the initial polarization:
for a large tipping angle $\theta=\pi/3$ as in the experiment, the
Leggett-Rice parameter $\gamma$ saturates at larger values than in the
case of a small tipping angle $\theta=\pi/30$.  The value to which
$\gamma$ saturates also depends on the strength of the gradient
$\alpha$, in contrast to the homogeneous case (see below).
Figure~\ref{fig:intdiff} also shows that the Leggett-Rice parameter
changes sign, reflecting the sign of the effective interaction between
quasiparticles which is attractive on the BCS side and repulsive on
the repulsive branch \cite{trotzky2015}.  The sign change occurs for
slightly positive values of $1/k_Fa$.

The center panel in Fig.~\ref{fig:intdiff} shows the bare transverse
diffusivity $D_0^\perp$.  Again, the trap diffusivity is significantly
lower than the homogeneous value (solid black line) at weak coupling
and agrees with experiment.  Once $\gamma$ is known, the bare
diffusivity $D_0^\perp$ is found from the fit of the analytical
solution \eqref{eq:lambert} to the magnetization decay in
Fig.~\ref{fig:ampphase}.

The bottom panel in Fig.~\ref{fig:intdiff} displays the ratio
\begin{align}
  \label{eq:lambda0}
  \lambda_0=-\frac{\hbar\gamma}{2mD_0^\perp} = \frac{Wn}{2m\alpha_\perp}.
\end{align}
It measures the strength of the effective interaction irrespective of
the scattering time $\tau_D$ and follows the sign change of $\gamma$
since $D_0^\perp > 0$.  Again, the trap calculation agrees with the
experimental data except in the instability region
$0 < 1/k_Fa \lesssim 1.3$ \cite{trotzky2015}, while the homogeneous
gas has more pronounced interaction effects.

Spin rotation in a trapped gas is determined qualitatively by the
ratio between different length scales: the trap size
$R_\text{TF} \approx 5\mu$m, the helix pitch
$\ell_\text{helix} = 2\pi/(\alpha t_e/2) \approx 4\mu$m for $t_e=2$\,ms,
and the mean free path $\ell_\text{mfp} = v_F \tau_D$, which ranges
from $0.5\mu$m at unitarity to $3.8\mu$m at $1/k_Fa=-2$.  In
simulations with a gradient $\alpha$ larger than the experimental
value, several helix pitches fit into the trap,
$R_\text{TF} \gg \ell_\text{helix}$, and the trap averaged diffusivity
and $\gamma$ are \emph{equal} to their homogeneous values as long as
also $\ell_\text{mfp} < \ell_\text{helix}$, but saturate for larger
$\ell_\text{mfp}$.  This is in marked contrast to longitudinal spin
diffusion, where a scaling factor of about $5$ was found to relate the
trap-averaged and homogeneous diffusivity $D_\parallel$
\cite{sommer2011a, bruun2011, *bruun2011spin, enss2012spin}.  For
weaker gradients where $\ell_\text{helix} > R_\text{TF}$, less than
one helix pitch fits into the trap and the homogeneous solution is
reached not even at unitarity where $R_\text{TF} \gg \ell_\text{mfp}$.

\begin{figure}[t]
  \centering
  \includegraphics[width=\linewidth]{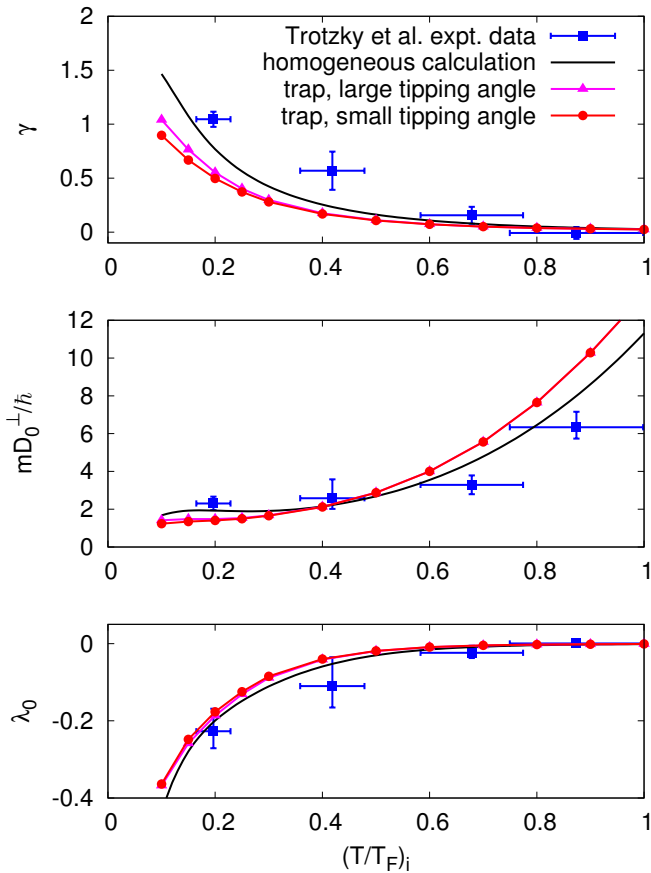}
  \caption{\label{fig:tempdiff}(Color online) Temperature effect on
    spin transport at unitarity: Leggett-Rice parameter $\gamma$, bare
    diffusivity $D_0^\perp$, and interaction parameter $\lambda_0$ as
    a function of the initial temperature $(T/T_F)_i$.  At unitarity
    the results for small ($\theta=\pi/30$; circles) and large
    ($\theta=\pi/3$; triangles) tipping angles differ very little;
    (blue) squares are experimental data \cite{trotzky2015}.}
\end{figure}
Finally, Fig.~\ref{fig:tempdiff} shows the temperature dependence of
the Leggett-Rice parameter at unitarity.  $\gamma$ decreases with
temperature, which is understood as follows: the two-body scattering
amplitude is purely imaginary at unitarity and would imply a vanishing
$\gamma \sim \Re \tmat$.  Hence, the observed finite value of $\gamma$
in the quantum degenerate regime $(T/T_F)_i < 1$ is a \emph{many-body}
effect due to medium scattering.  The presence of the medium enhances
both the dissipative and the reactive effects of scattering at low
temperatures, and more so in the homogeneous gas than in the trap, where
only the core is strongly interacting.

Similarly, the diffusivity reaches values of
$D_0^\perp \simeq 2\hbar/m$ in the quantum degenerate regime (center
panel), in agreement with experiment, while the ratio $\lambda_0$
becomes large only at the lowest temperatures (lower panel).


\section{Conclusion}
\label{sec:con}

To summarize, a spin-echo sequence in a trapped Fermi gas is modeled
by a nonlinear and complex diffusion equation for the transverse
magnetization.  The spin evolution exhibits the Leggett-Rice effect of
strength $\gamma=\mu n$, which appears to saturate for weak coupling,
and a bare diffusivity much lower than expected for a homogeneous gas
of the same temperature and interaction strength.  These results are
obtained without any fit parameters and agree very well with the
weak-coupling data measured recently for a trapped gas of ultracold
fermionic $^{40}$K atoms \cite{trotzky2015}.  The present calculation
provides an intuitive interpretation of the observed saturation of
$\gamma$: while the spin helix in the homogeneous gas is completely
unwound at the echo time $t_e$, the trapped gas remains partially in a
helical state, with the average transverse magnetization $A(t_e)$
strongly reduced and a smaller phase shift $\phi(t_e)$.  At weak
coupling $1/\abs{k_Fa} \gtrsim 2$ the kinetic theory employed in this
study is well controlled and includes the relevant interaction
effects.

At strong coupling $1/k_Fa=0$ (unitarity), the Leggett-Rice effect is
absent at the two-body level and arises only due to many-body medium
scattering.  The trap calculation at strong coupling agrees with the
experimental results qualitatively; the remaining differences may be
due to reheating in a demagnetized Fermi gas \cite{bardon2014,
  trotzky2015} or due to interaction corrections to the equation of
state which were not included in this study; they remain a topic for
future work.

Ultracold Fermi gases are thus ideal systems to study the interaction
and temperature dependence of the Leggett-Rice effect and the spin
transport coefficients.  The spin dynamics of the trapped gas
(Fig.~\ref{fig:profiles}) is much more complex than in the homogeneous
case and requires a numerical solution.  The spin evolution simplifies
near unitarity if $R_\text{TF} > \ell_\text{helix} > \ell_\text{mfp}$,
in which case the homogeneous solution is recovered without any
trap-related scaling factors.  The decay of the trap-averaged
magnetization can be analyzed by the methods developed for the
homogeneous case: the magnetization decay fits the homogeneous
solution surprisingly well (Fig.~\ref{fig:ampphase}).  This study
shows that the extracted transport coefficients for the trapped gas
can differ markedly, especially at weak coupling, from those of the
corresponding homogeneous system (Fig.~\ref{fig:intdiff}).
Conveniently for the interpretation, the sign of $\gamma$, which
reveals the repulsive or attractive character of the effective
interaction, does not change with the trap average.

The author wishes to thank E. Taylor, J. Thywissen, S. Trotzky, and
S. Zhang for stimulating discussions.

\bibliography{all}

\end{document}